\documentclass{sigchi-ext}
\usepackage[T1]{fontenc}
\usepackage{textcomp}
\usepackage[scaled=.92]{helvet} % for proper fonts
\usepackage{graphicx} % for EPS use the graphics package instead
\usepackage{balance}  % for useful for balancing the last columns
\usepackage{booktabs} % for pretty table rules
\usepackage{ccicons}  % for Creative Commons citation icons
\usepackage{ragged2e} % for tighter hyphenation

\def\plaintitle{Human-Drone Interactions in Complex Mission Environments} 
\def\emptyauthor{}
\def\plainkeywords{Human-drone collaboration, emergency response}

\title{Human-Drone Interactions with Semi-Autonomous Cohorts of Collaborating Drones}

\numberofauthors{2}
\author{%
  \alignauthor{%
    \textbf{Jane Cleland-Huang}\\ %First author
    \affaddr{University of Notre Dame} \\
    \affaddr{Notre Dame, IN 46556, USA} \\
    \email{JaneHuang@nd.edu}\vspace{16pt} }
    \alignauthor{%
    \textbf{Ankit Agrawal}\\ %Second author  ** THe template requires column first!! Its weird.
    \affaddr{University of Notre Dame}\\
    \affaddr{Notre Dame, IN 46556, USA}\\
    \email{aagrawa2@nd.edu}} 
    \alignauthor{
  }}

\definecolor{linkColor}{RGB}{6,125,233}
\hypersetup{%
  pdftitle={\plaintitle},
  pdfauthor={\emptyauthor},
  pdfkeywords={\plainkeywords},
  bookmarksnumbered,
  pdfstartview={FitH},
  colorlinks,
  citecolor=black,
  filecolor=black,
  linkcolor=black,
  urlcolor=linkColor,
  breaklinks=true,
}

\begin{document}

\copyrightinfo{This paper is published under the Creative Commons Attribution 4.0 International (CC-BY 4.0) license. Authors reserve their rights to disseminate the work on their personal and corporate Web sites with the appropriate attribution.\\
\emph{Interdisciplinary Workshop on Human-Drone Interaction (iHDI 2020)\\
CHI ’20 Extended Abstracts, 26 April 2020, Honolulu, HI, US}\\
\copyright~Creative Commons CC-BY 4.0 License.}

%\copyrightinfo{\acmcopyright}

\maketitle

% Uncomment to disable hyphenation (not recommended)
% https://twitter.com/anjirokhan/status/546046683331973120
\RaggedRight{} 

% Do not change the page size or page settings.
\begin{abstract}
Research in human-drone interactions has primarily focused on cases in which a person interacts with a single drone as an active controller, recipient of information, or a social companion; or cases in which an individual, or a team of operators interacts with a swarm of drones as they perform some coordinated flight patterns. In this position paper we explore a third scenario in which multiple humans and drones collaborate in an emergency response scenario. We discuss different types of interactions, and draw examples from current DroneResponse project.
\end{abstract}


% ACM Classfication



\section{Introduction}
Small Unmanned Aerial Systems, which we refer to here as drones, can be effectively deployed to support emergency responders for diverse scenarios such as search-and-rescue, accident surveillance, and flood inspections.  Currently, emergency responders tend to operate drones manually or using off-the-shelf applications that allow them to preprogram sets of waypoints.  However, equipping drones with onboard intelligence allows them to perform tasks autonomously and to contribute more fully to the emergency response.

\begin{marginfigure}[0pc]
  \begin{minipage}{\marginparwidth}
    \centering
    \includegraphics[width=1\marginparwidth]{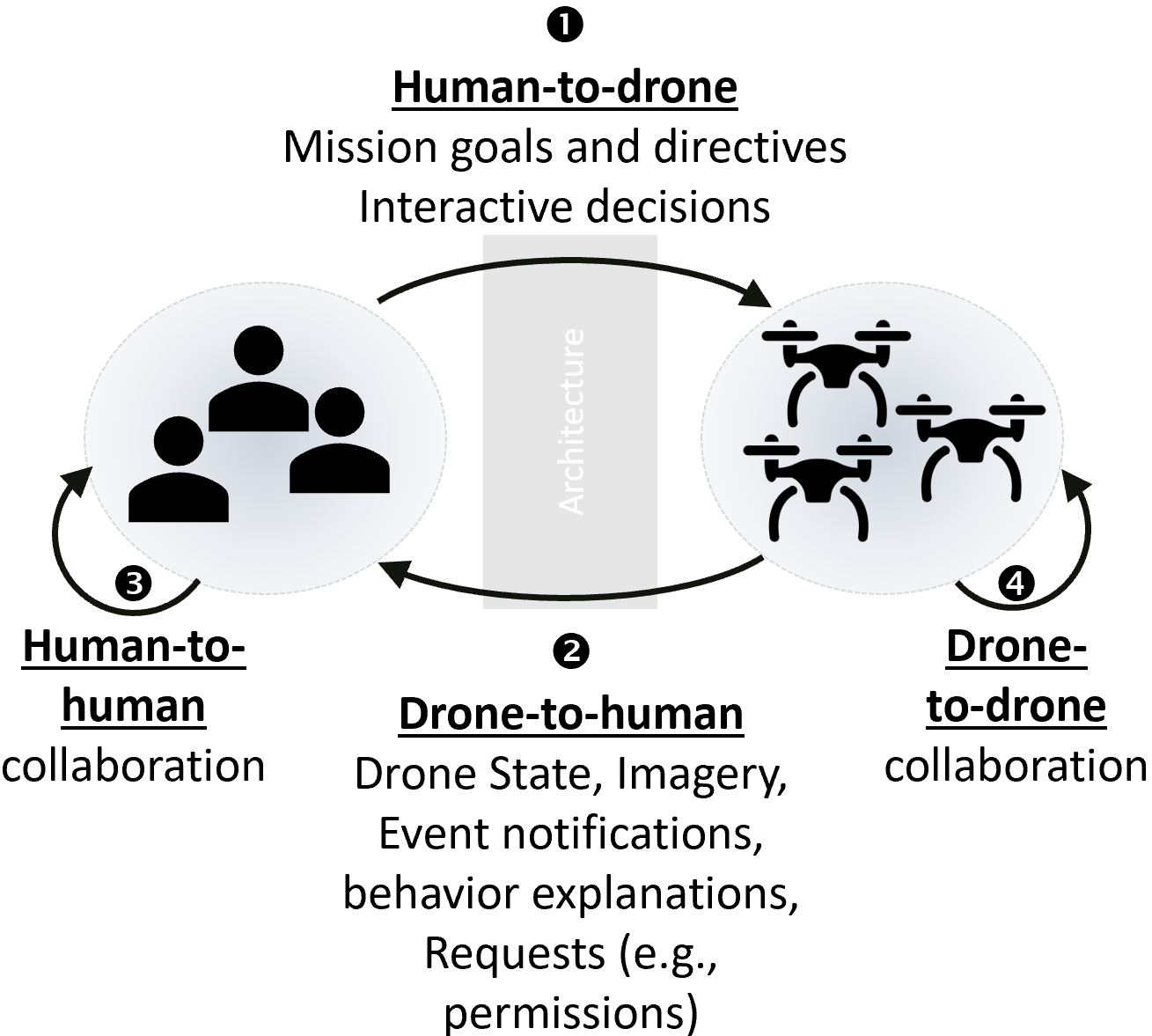}
    \caption{Humans and drones interact in many different ways.  \newline }~\label{fig:HDI}
  \end{minipage}
\end{marginfigure}

In our DroneResponse project we are designing and developing a system to deploy and coordinate the efforts of multiple semi-autonomous drones for use in emergency situations \cite{chi2020,DBLP:conf/re/Cleland-HuangV18}.  Our vision is for humans and drones to work closely together as part of a complex mission -- for example to monitor air quality following a chemical explosion, to perform search and rescue, to deliver medical supplies, or to support firefighters during structural fires.  As depicted in Figure \ref{fig:HDI}, there are several facets to human-drone interaction in such scenarios. Humans need to communicate mission goals and directives to groups of drones as well as to individuals, while drones need to keep humans informed of their current state and progress, and at times, need to seek permission or guidance to perform specific tasks. In addition, both humans and drones need to communicate between themselves (i.e., drone-to-drone and human-to-human) to coordinate their activities. 

\section{An Interaction Example}

\begin{figure*}[!t]
\centering
	\includegraphics[width=.8\textwidth]{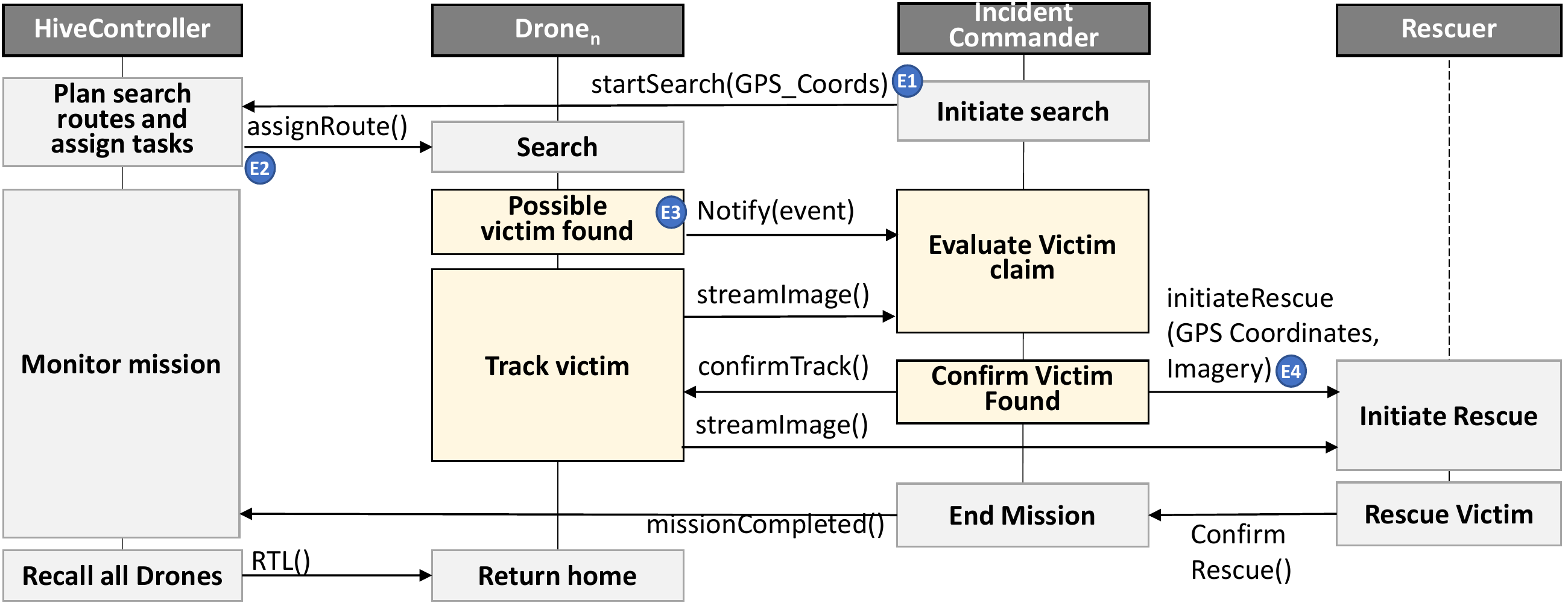}
	\caption{A Sequence Diagram showing the interactions between humans and drones for the scenario in which a drone searches for, and detects, a potential drowning victim.}
	\label{fig:sequence}
\end{figure*}

We provide examples of such interactions in the sequence diagram depicted in Figure \ref{fig:sequence}.  The \emph{Incident Commander} first defines a search area and sends a request to the \emph{hive controller} to start the search (E1).  This is an example of human-to-drone interactions (H2D).  The hive-controller then creates a search plan and assigns search routes to drones (E2).  The coordination between drones represents drone-to-drone (D2D) interaction. In the modeled sequence of actions, \emph{Drone$_n$} detects a potential drowning victim. It then notifies the human incident commander and starts streaming annotated video to the ground (E3), thereby illustrating drone-to-human communication (D2H). The part of the sequence diagram highlighted in yellow provides an example of a more complex bi-directional human-drone conversation in which the drone uses its sensing (image detection) abilities to detect a victim. It then autonomously switches to \emph{track-victim} mode, raises a \emph{victim-found} alert, and streams annotated video to the incident commander. Finally, the incident commander uses the information relayed by the drone, to confirm the victim sighting and to push information from the drone to a physical rescue team (E4). This final step is an example of human-to-human (H2H) interaction, triggered by the initial D2H exchange. This sequence of events illustrates the complex socio-technical aspects of emergent multi-user, multi-drone interaction spaces.

\section{Drone-to-Human Communication (D2H)}
There are numerous challenges that must be addressed in order to achieve efficient human-drone collaboration. In our concurrently published work \cite{chi2020}, we have focused on designing a user interface that enables situational-awareness (SA) \cite{endsley2012} for human first-responders. SA involves perception (i.e., recognizing and monitoring), comprehension (i.e., interpreting and synthesizing information), and projection (i.e., understanding the situation, projecting future outcomes) so that a user can make effective and actionable decisions. The key to designing D2H communication is identifying information that is needed by different user roles within specific contexts.  As an example, a drone may be ascribed the ability to autonomously decide its speed and altitude during a search.  If visibility is good, the drone might fly higher and faster in order to cover the search area more quickly, while still returning accurate results.  On the other hand, if visibility is lower, the drone might need to fly lower and slower, and adapt its flight plan to compensate for a reduced field of view. In this scenario, the operator needs visual cues and awareness of why a drone behaves as it does.  As an outcome of a four month co-design process with our local fire department, we identified two design strategies to address this specific scenario.  First, we designed our DroneResponse GUI to depict any environmental factors that were likely to impact drone behavior -- for example, low visibility, high-winds, or prohibited airspace.  Secondly, we made the drones \emph{explain themselves} upon demand by describing their current strategies and permissions.  In the case of searching for a victim in inclement weather, the drone might explain ``flying lower than normal at 10 meters due to low visibility'' or ``searching river banks at a greater distance than normal due to high wind gusts and moving branches.''  We report outcomes from our co-design experience, especially with respect to D2H interactions and achieving situational awareness in our related paper \cite{chi2020}. 
\begin{marginfigure}[-20pc]
  \begin{minipage}{\marginparwidth}
    \centering
    \includegraphics[width=1\marginparwidth]{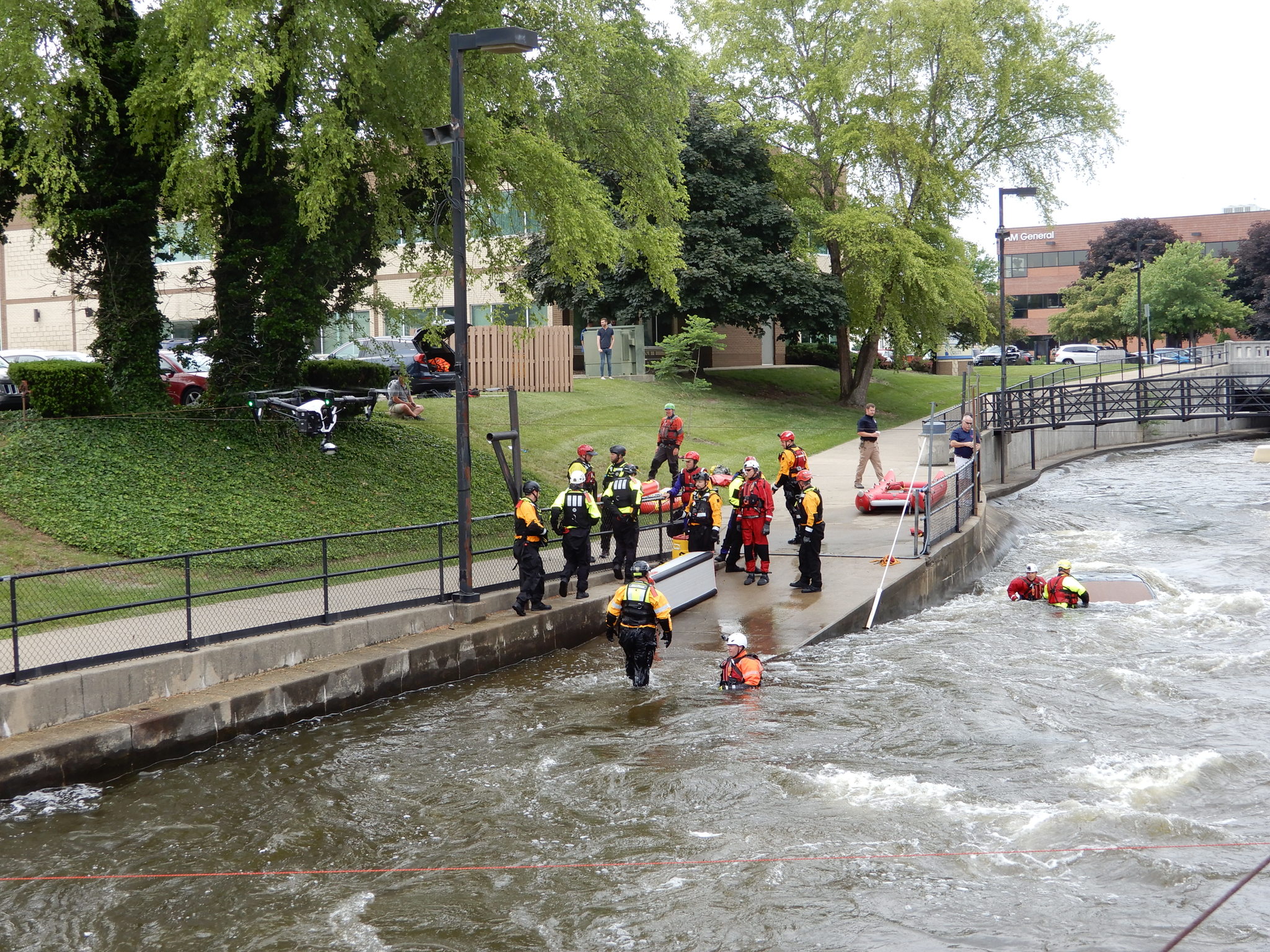}
    \caption{Firefighters in South Bend manually operate drones}~\label{fig:HDI1}
  \end{minipage}
    \begin{minipage}{\marginparwidth}
    \centering
    \includegraphics[width=1\marginparwidth]{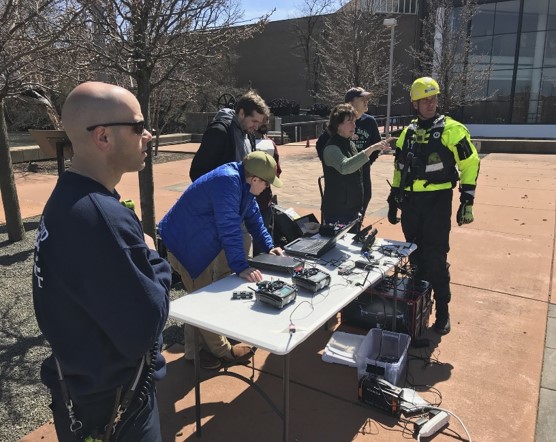}
    \caption{Notre Dame researchers and firefighters interact with multiple drones using Dronology} \cite{DBLP:conf/icse/Cleland-HuangVB18}~\label{fig:HDI2}
  \end{minipage}
    \begin{minipage}{\marginparwidth}
    \centering
    \includegraphics[width=1\marginparwidth]{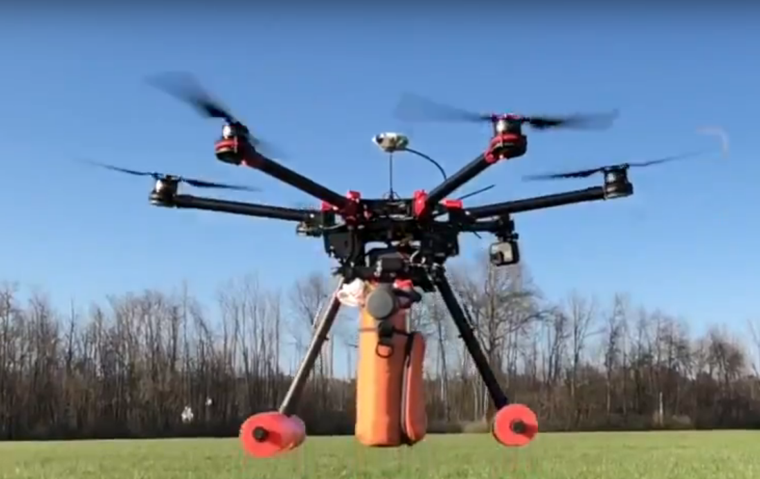}
    \caption{Defibrillator delivery by drone requires remote UI }~\label{fig:HDI3}
  \end{minipage}
\end{marginfigure}

\section{Human to Drone Mission Directives}
Achieving effective H2D communication is challenging in systems with multiple drones and complex missions. Several research groups have explored ways to specify drone missions using formal commands, often embedded in domain specific languages \cite{10.1145/3357766.3359535}. However, it is infeasible for emergency responders to write such mission specifications under stressful time-constraints of a life-and-death response. A user interface is therefore needed that enables quick mission planning and configuration and which supports high-level directives addressed to the cohort of drones, as well as specific directives addressed to individual drones.

\begin{figure}[!t]
    \centering
    \includegraphics[width=1\columnwidth]{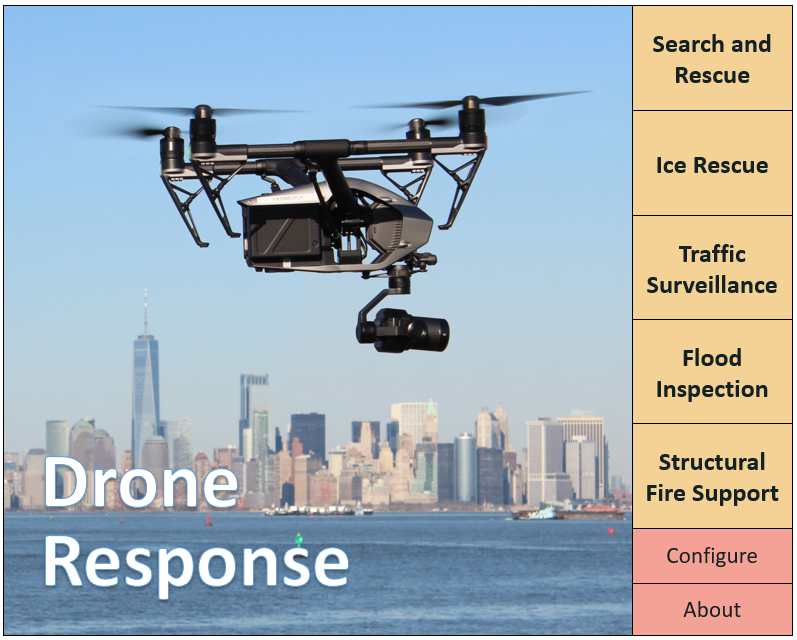}
    \caption{A user initially selects a predefined mission type which include mission objectives and task specifications. \newline }~\label{fig:GUI-splash}
\end{figure}

Researchers have previously explored diverse solutions for issuing commands to drones, such as the use of gestures and voice commands \cite{DBLP:journals/interactions/Funk18,hdi-survey} or airplane-like cockpits for controlling large military-style drones \cite{dronecockpit}. We have prototyped the use of gestures and voice commands; however, they have several shortcomings that inhibit their use in emergency response scenarios. Voice commands, while appealing, are impractical due to the noise inherent to a rescue scene.  This includes sirens, constant radio-chatter, and now the additional noise of drone motors. Gestures are similarly impractical.  They have been shown to work effectively in controlled near-distance environments, which is far from the case for an emergency response scenario \cite{10.1145/2750858.2805823}. Furthermore, they introduce significant room for error, which is unacceptable in an emergency response environment, where mistakes could cost the lives of both the victims and the rescuers. Domain experts, collaborating in the co-design of DroneResponse soundly ruled out either of these approaches \cite{chi2020}. We therefore have opted to create a GUI-based solution for HD2 commands for emergency response missions.
\begin{marginfigure}[0pc]
  \begin{minipage}{\marginparwidth}
    \centering
    \includegraphics[width=1\marginparwidth]{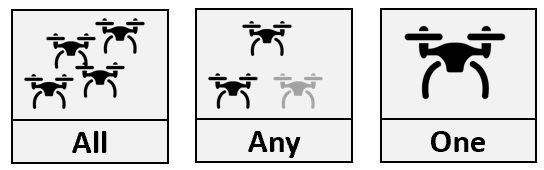}
   \caption{Mission commands can be addressed to specific groups of drones.\vspace{4pt}\newline}
   \label{fig:UI1}
   \vspace{4pt}
   \end{minipage}
  \begin{minipage}{\marginparwidth}
    \centering 
    \includegraphics[width=1\marginparwidth]{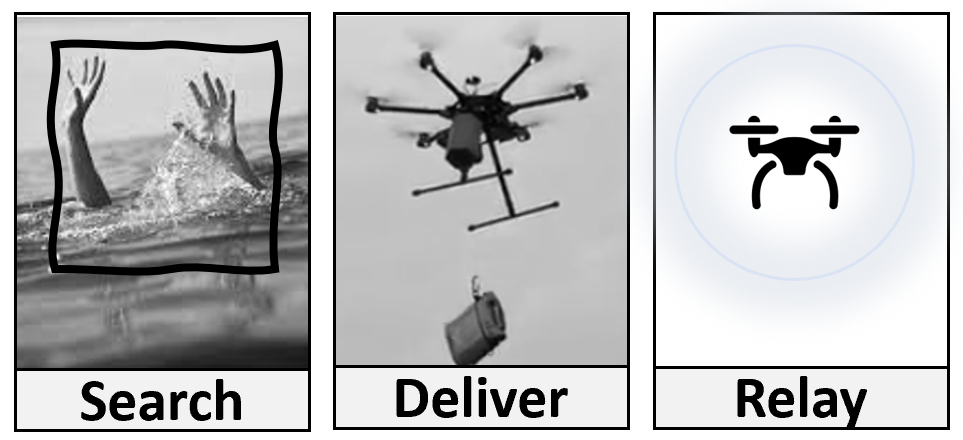}
   \caption{High-level commands, such as search, deliver, or relay, are domain specific.  These ones target river search and rescue.\vspace{4pt}\newline}
      \label{fig:UI2}
   \end{minipage}
   \begin{minipage}{\marginparwidth}
    \centering
       \vspace{4pt}
   \includegraphics[width=1\marginparwidth]{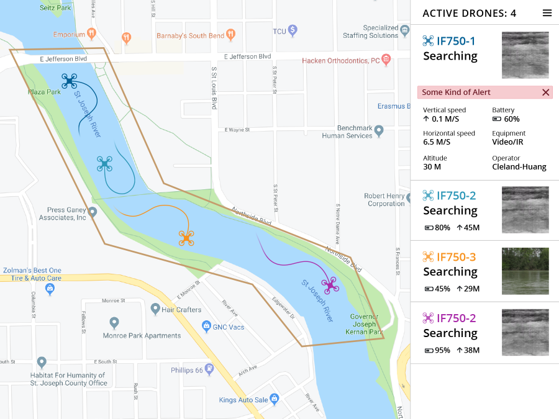}
    \caption{The user can mark an area on the map to define either a region or a point of interest to be targeted by the H2D directive. \newline }~\label{fig:GUI-interface}
       \label{fig:UI3}
  \end{minipage}
\end{marginfigure}
In our GUI, which is currently under development, users can initially select a mission type from a high-level list of missions as depicted in Figure \ref{fig:GUI-splash}. They then perform a series of configurations such as marking a search area.  Each predefined mission type will have a corresponding underlying mission plan with configuration points.  This plan is sufficient for allowing the mission to proceed through a series of predefined stages and tasks (e.g., search, track, return home).  However, users will also need to configure or tweak the mission dynamically as it evolves, by providing additional directives.  

Each of these directives must specify \emph{who}, \emph{what}, \emph{where}, and \emph{how} a task is to be accomplished. `Who' refers to whether the command is addressed to the entire cohort or an individual drone.  In the case of the cohort, then the \emph{hive coordination} is empowered to autonomously figure out which drones are best fit to respond.  'Who' could also be specified with constraints such as \emph{3 drones} or \emph{drones with thermal cameras onboard}. Finally `who' could be addressed to a specific drone if it were the case, that the Incident Commander wished to assign a task to a specific drone. This is more risky, as the selected drone might be unfit for service (e.g., due to low battery or current critical service). `What' refers to the specific task to be completed -- for example reconnaissance, delivery, or serving as a communication relay if drones are communicating using onboard communication channels such as ad-hoc wifi. `Where' refers to a region or point of interest defined  by GPS coordinates. For example, in the case of reconnaissance, the user might need to direct the drone to a certain part of a wooded river bank where somebody has sighted a piece of clothing; while in the case of establishing a communication relay, the user could either specify coordinates or allow it to dynamically position itself so as to optimize communication between all drones.  Finally, `how' enables specific directives for how the task is to be completed. In some cases, the drones could be given significant autonomy to complete well-defined tasks, while in other cases more specific guidelines might be required. We are currently working closely with several emergency response organizations to better discover their needs and to formally model diverse mission plans.
\begin{marginfigure}[0pc]
  \begin{minipage}{\marginparwidth}
    \centering
   	\includegraphics[width=0.95\marginparwidth]{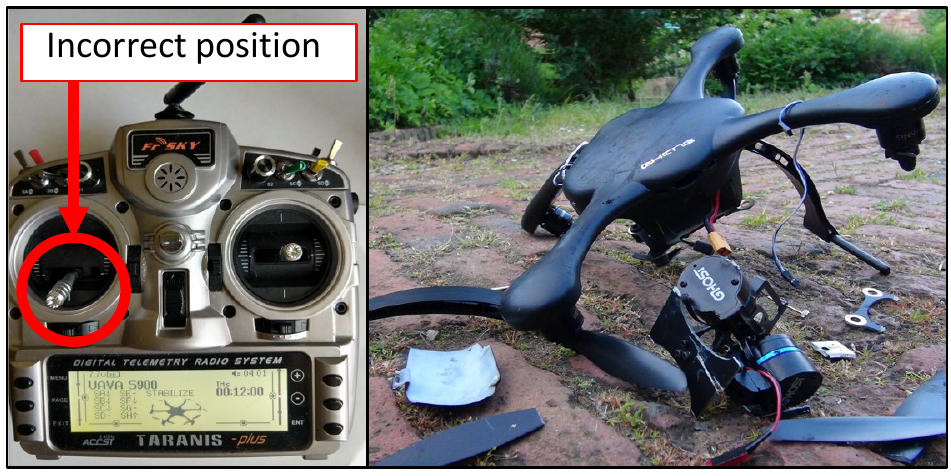}%Fault.pdf}
\caption{The throttle on the physical user interface was incorrectly positioned during a transition from software-controlled flight to manually operated flight, causing the UAV to crash.}
		\label{fig:failure}
    \end{minipage}
\end{marginfigure}
The GUI therefore provides a human-facing interface to an underlying mission plan specified using a more formal approach such as belief-desire-intent \cite{rao1995bdi}.  Drones are able to interpret the more formal specification. In our initial prototype we are experimenting with a flow-chart of buttons to enable humans to configure mission directives in a known space of options.  These are depicted in Figures \ref{fig:UI1}-\ref{fig:UI3}.

\section{GUI versus Physical Devices}
The discussion in this position paper has focused almost entirely on human-drone interfaces based on the use of graphical interfaces; however, drones can also be controlled using physical hand-held devices. In a multi-drone scenario, humans might need to switch between graphical and physical interfaces for several reasons including taking manual control of a malfunctioning drone or temporarily using manual controls for a specific task that is currently beyond the capabilities of a drone to perform autonomously. Our prior work has shown that misalignment of GUI's and physical controllers can easily lead to accidents \cite{DBLP:conf/re/Cleland-HuangV18} (see Figure \ref{fig:failure}.  For example, when control is passed from a computer to a handheld device, the physical switches on the hand-held controller must be set to stable 'flightmode' positions, otherwise accidents, including crashlandings, could occur. Furthermore, when humans take-over control of remote drones, it can be exceedingly difficult to figure out which direction the drone is facing. Commands are interpreted relative to the drones position, which means that issuing a `forward' command would cause the drone to fly forward, but without clear orientation from the remote-pilot's perception, that could actually be in any direction. A simple design solution might be to provide a feature to autonomously reorient the drone with respect to the remote pilot so that physical and GUI controls become aligned relative to the drone's and pilot's positions. Given this reorientation, a  \emph{forward} command would then consistently send the drone away from the pilot, and a \emph{moveRight} command would make it move right. For deployment in emergency situations, more though should be invested in the use of both graphical and physical interfaces,  the interactions between them, and transitions of control across devices and between different operators.

\section{Conclusion}
This position paper has presented an informal framework for considering human-drone interactions along the dimensions of H2D, D2H, D2D, and H2H communication in multi-user, multi-drone environments where drones are permitted to operate with some degree of autonomy. We have described some of the challenges we are facing in the design of DroneResponse and some initial ideas for addressing those challenges. Our prior work \cite{chi2020} has focused primarily on the D2H challenge of supporting situational awareness, while our ongoing work focuses on providing a meaningful interface for more complex bidirectional human and drone interactions.

\section{Acknowledgements}
The work described in this position paper has primarily been funded by the National Science Foundation under grants CNS-1737496, CNS-1931962, CCF-1647342, and CCF-1741781. We also thank the Firefighters of South Bend for closely collaborating on the DroneResponse project.

\balance{} 

\bibliographystyle{SIGCHI-Reference-Format}
\bibliography{drone}

\end{document}